# A Semantic Grid Oriented to E-Tourism

Xiao Ming Zhang

College of Computer and Communication, Hunan University, 410082 Changsha, China
School of Hospitality & Tourism Management, Florida International University, 33181 Miami, USA

zhangxm19712003@yahoo.com.cn

**Abstract.** With increasing complexity of tourism business models and tasks, there is a clear need of the next generation e-Tourism infrastructure to support flexible automation, integration, computation, storage, and collaboration. Currently several enabling technologies such as semantic Web, Web service, agent and grid computing have been applied in the different e-Tourism applications, however there is no a unified framework to be able to integrate all of them. So this paper presents a promising e-Tourism framework based on emerging semantic grid, in which a number of key design issues are discussed including architecture, ontologies structure, semantic reconciliation, service and resource discovery, role based authorization and intelligent agent. The paper finally provides the implementation of the framework.

## 1    Introduction

Tourism has become the world's largest industry, composing of numerous enterprises such as airlines, hoteliers, car rentals, leisure suppliers, and travel agencies. The World Tourism Organization predicts that by 2020 tourist arrivals around the world would increase over 200% [1]. In this huge industry, e-Tourism representing almost 40% of all global e-Commerce [2] is facing a need of the next generation infrastructure to support more innovative and sophisticated tasks like dynamic packaging, travel planning, price comparison, travel route design, and multimedia based marketing and promotion. Fig. 1 shows an e-Tourism task scenario which can illustrate several use cases and aid in capturing feature requirements for the future e-Tourism infrastructure.

- **Use case 1:** A travel agency receives a tourist's request of travel plan for the holiday (step 1), so it enters a VO (Virtual Organization) which is able to organize necessary, rich and authorized tourism services and resources in a cross-institutional way. Then the travel agency forwards tourists' travel plan task to an agent (step 2). Next, the agent coordinates other agents and assigns them smaller branch tasks (step 3). In fig. 1, one branch task is to design travel route which performs complex calculations based on GIS and Traffic database, and the other two branch tasks are queries of related air fare and hotel room rates through GDS (Global Distribution Framework) and CRM (Central Reservation Management). Finally the first agent processes the results from its partners and returns travel plans to the travel agency (step 4) and the tourist (step 5).

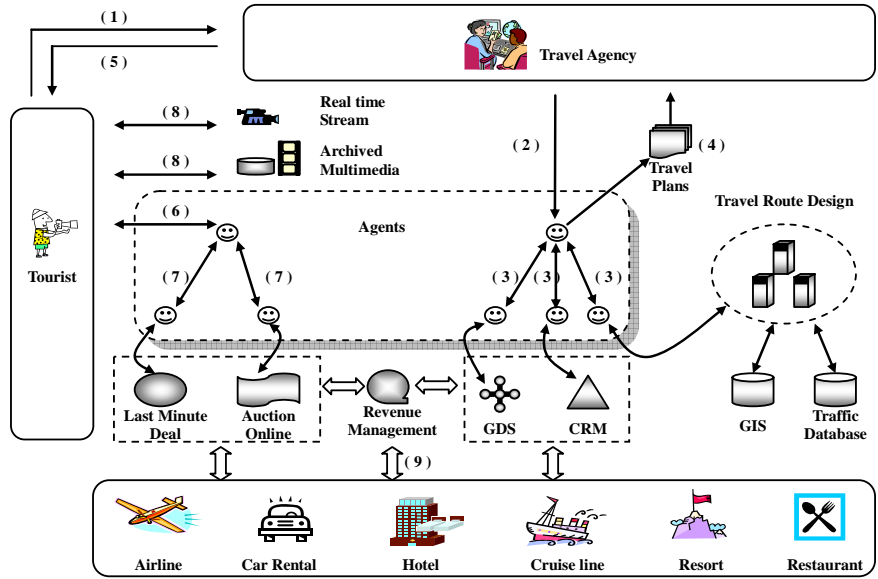

**Fig. 1.** An e-Tourism Task Scenario

- **Use case 2**: The authorized tourist directly enters the VO and search required services. In this situation, the system needs to trace and analyze user's preferences and requirements, then compare them with services' functionalities. In fig. 1, the system discovers and returns Last Minute Deal and Auction Online services to the tourist who interests in the best room rates of hotel (step 6, 7).
- **Use case 3**: To promote the products or services, some services providers conduct a multimedia marketing, allowing tourists watch large amount of archived photos and live streaming video on line (step 8).
- **Use case 4**: The services are able to access other services and resources owned by different service providers seamlessly. For example, some advanced service providers prefer to use Revenue Management service (step 9), which is able to interact with other sales online systems (e.g. GDS) and maximize business profit through differentiating the prices of different products in those online systems.

From the above use cases, several feature requirements for the future e-Tourism infrastructure can be identified:

- Personalization and intelligent agent. The individual intention and tasks should be able to be personalized and represented in intelligent agents who can interact with other agents and objects, perform various tasks, deal with events timely and seek solutions to the targets autonomously.
- Knowledge service. It includes discovering resource and services through inference, annotating the contributed resource, combining content from multiple sources, and scheduling the agents and workflow to the tasks.

- High degree of automation. It refers to describe the services, resources and even workflow in a machine understandable manner, and thus different tasks like travel planning could be performed in an automatic way with a good performance, without or with very limited human support.
- Seamless integration. Various resources, services and relevant users should be integrated in a uniform and seamless manner, realizing the dynamically and semantically enhanced information processing.
- Computations. The distributed, extensible and transparent computation resource is required by the computing intensive services like travel route design.
- Storages. The system should be able to store and process potentially huge volumes of multimedia content, user information and business data in a timely and efficient fashion.
- VO and collaborations. VO is required to make users share resources, access services and collaborate in a cross-institutional way.

To meet the above requirements, currently several enabling technologies like semantic Web [3], Web services, grid and agent have been adopted in the different e-Tourism applications, however there is no an unified framework to be able to seamlessly integrate all of them. Inspired by UK e-Science program, the paper focuses on the solution from a promising infrastructure, semantic grid [4][5]. Semantic grid is able to provide an Internet centered interconnection environment that effectively organize, share, cluster, fuse, and manage globally distributed versatile resources based on the interconnection semantics [6]. In this paper, we research an e-Tourism framework based on semantic grid, which conforms to the S-OGSA architecture [7] and enhanced by several customizations and extensions such as the ontologies, intelligent agent, unified service and resource discovery, etc.

The remainder of this paper is structured in the following manner. Section 2 provides framework overview. Section 3 discusses several design issues including architecture, ontologies structure, agent, etc. Section 4 presents the implementation. Section 5 introduces the related work and section 6 gives out the conclusion.

## 2  Framework Overview

The e-Tourism framework proposed in this paper is designed based on semantic grid, which conforms to three layered views: service view, content view and technological view (fig. 2).

Firstly, from the service view, the framework takes the notion of service-oriented, in which any users including tourist, travel agency, hotel, restaurant, airline and resort should be considered as either service consumer or service provider. VOs organize services and relevant resources and users in a virtual administrative domain, allowing the access of services in a cross-institutional way.

Secondly, content view refers to the objects that the framework can process from simple data to meaningful information, then to higher abstract knowledge. This view has been widely accepted since it is presented by David, Nicholas and Nigel [8]. In this view, the data is concerned with the way how it is obtained, shipped, processed and transmitted by services, resources and even some special equipments like camera;

then, the information is the data equipped with meaning, which is related to the way how it is represented, annotated, achieved, shared and maintained. For example, the number can be annotated as distance between a hotel and an airport; finally, the knowledge is the information aiding users to achieve their particular goals, which is concerned with the way how it is acquired, inferred, retrieved, published and maintained. For instance, a piece of knowledge can be stated like this: 20 minutes driving is needed from the airport to a hotel if the distance is about 20 miles.

Thirdly, technological view considers the state of the art technological components which implement the e-Tourism semantic grid. For instance, grid supports VO, computation, storage and collaboration management; agent and workflow facilities provide personalization and intelligent automation; semantic Web and Web services enable the seamless integration and knowledge inferring.

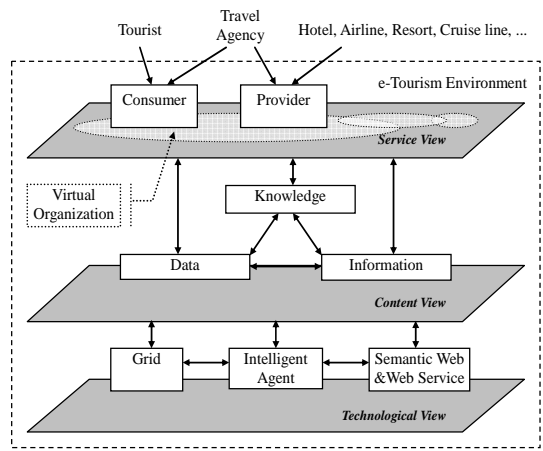

**Fig. 2.** Overview of E-Tourism Framework

## 3 Design Issues

In this section, some detailed design issues are discussed mainly focusing on e-Tourism specific requirements, which include framework architecture, ontologies structure, semantic reconciliation, service and resource discovery, role based authorization and intelligent agent.

### 3.1 Architecture

To avoid the reinvention of the wheel, our framework takes S-OGSA as the start point for architecture design, which is widely considered as the first reference architecture for semantic grid by extending the current OGSA (Open Grid Services Architecture) [9] with semantics. S-OGSA integrates the grid, semantic Web and semantic Services together, supports many important knowledge services. For example, ontology and reasoning services are designed for the conceptual models of representing knowledge, while metadata and annotation services are invented to implement semantic binding for different types of information sources, like documents, databases, provenance information, credentials, etc. However, as a basic and common architecture, S-OGSA can't be directly applied to all different applications, so according to the specific requirements of e-Tourism, we extend or enhance S-OGSA as followings:

- Extending ontologies structure with tourism domain requirements
- Introducing semantic reconciliation to solve the interoperability of ontologies
- Unifying the service and resource discovery
- Establishing role based authorization
- Integrating intelligent agent facility

### 3.2 Ontologies Structure

In the e-Tourism semantic grid, ontologies are the fundamental blocks to capture the expressive power of modeling and reasoning with knowledge [8].

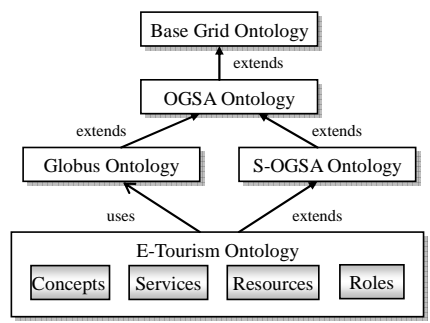

**Fig. 3.** Ontologies Structure

From the ontologies structure illustrated in Fig. 3, e-Tourism ontology is obtained through extending the S-OGSA ontology and using part of the Globus [10] ontology. To meet the specific requirements in tourism domain, e-Tourism ontology is composed of four parts: Concepts ontology provides basic definitions and standard terms based on WTO (World Tourism Organization) Thesaurus [11], which is an international standardization and normalization guide to tourism terminology;

Roles ontology contains descriptions of user roles like tourist, travel agency and service provider; Resources ontology states the capability of the hardware, software and communication resources which support the services, for example, CPU performance and network bandwidth; Services ontology defines the uniform service interfaces and functionalities conforming to the OTA specification [11], which includes air services, cruise services, destination services, dynamic package services, golf services, hotel services, insurance service, loyalty services, tour services and vehicle services. Fig. 4 gives out hierarchical structure of OTA-Compliant services ontology and an example interface description of TourSearchRQService.

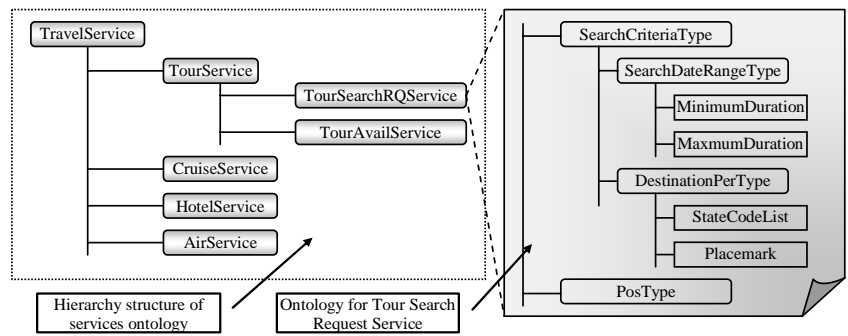

**Fig. 4.** OTA-Compliant E-Tourism Services Ontology

### 3.3 Semantic Reconciliation

Although e-Tourism ontology is helpful in establishing the new semantically interoperable services, it's impractical to reinvent all legend services and make them complaint to a global ontology in the tourism domain. Some different ontologies mentioned at the related work of this document have been adopted by several legend tourism systems or services, so there is a need of semantic reconciliation which typically implemented through ontology mediation or mapping in some projects[12][13][14].

In this framework, semantic reconciliation is solved mainly by stub service [15][16] and ontology service. Stub service encapsulates the details of translating between OTA-Compliant messages and Provider-Specific messages. In translation, stub services collaborate with ontology service which supports ontologies mapping besides the storage capability. Upon completing transformation of messages, stub services forward them to the target Provider-Specific services. In fig. 5, an OTA-Compliant request to TourSearchRQService is converted into Amadeus-specific and Sabre-specific requests in the different stub services through the ontology mapping.

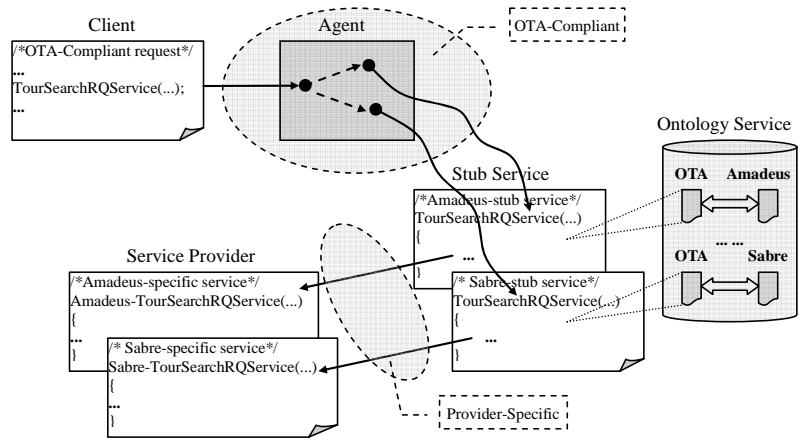

**Fig. 5.** Semantic Reconciliation

### 3.4 Unified Service and Resource Discovery

Service and resource discovery is critical to the dynamic and heterogeneous e-Tourism, however current discovery approaches such as UDDI and MDS [17] can't provide sufficient expressiveness and efficient matchmaking, so several researches focus on extending the existed discovery mechanism by semantic technologies [18]. For example, myGrid [19][20] implements the service discovery by attaching the semantic information to entities in the UDDI and WSDL models [21], and S-MDS (Semantic Monitoring and Discovery system) conducts discovery of grid resource through extending the Globus MDS with enhancement of semantics [18].

Based on the above achievements, our discovery solution strives to unify semantic discovery of service and resource together, because the service discovery in the grid is not only relied on the requirements of functionalities and features, but also the availability and performance of associated underlying resources [22]. As illustrated in fig. 6, the information of service and its respective stub service is stored in UDDI; the relied resource information with property of related service name and UDDI index is stored in MDS; the association among capability, service and its stub service is described in OWL-SR language [22] and stored in Mediator. At runtime clients or agents send semantic query to Mediator, for example "find a tour search service on a resource with CPU utilization less than 10%". Then Mediator infers a list of services which meet the service capability description through contacting ontology service and reasoning service. Next, Mediator infers and sends a query to the MDS based on this service list and the resource requirement. Upon receipt of resources from MDS, Mediator is able to narrow down the service list by kicking out the services without required resources. Finally Mediator forms a list composing of the stub services corresponding to the previously narrowed service list, and returns it to the clients or agents. This last step is critical to the system integration in the heterogeneous e-Tourism, because the OTA-Compliant interfaces provided by stub services are able to hide the interface difference of specific providers, which in turn simplifies and unifies the client and agent programming.

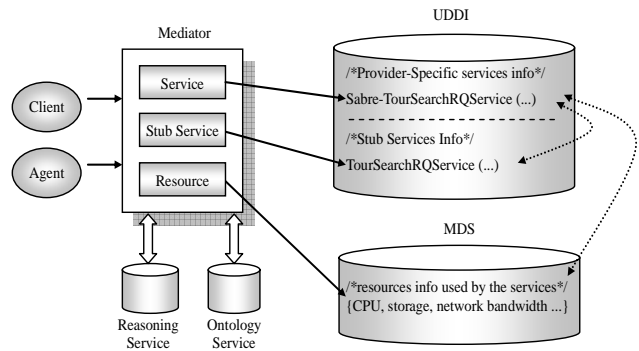

**Fig. 6.** Unified Service and Resource Discovery

### 3.5 Role Based Authorization

VO provides an effective way for cross-institutional services and resources accessing in the e-Tourism system. However, the authorization of the users from the different administrative domain cannot be pre-determined statically until at runtime. So we make a role based access control policy that conforms to the OGSA-AuthZ framework and is able to determine the users' eligibility dynamically [23] [24]. Under this policy, users' roles can be inferred from their properties during runtime. For example, EcnomicTourist role is assigned to a tourist whose consuming points in the past 12 months are below 5000, and he can be upgrades to VipTourist role when his consuming points in the past 12 months reach 5000 or above.

The fig. 7 illustrates the authorization process for a tourist to access RouteDesign service which is only authorized to VipTourist users. Initially Authorization service maintains an access control list based on roles. In step 1, a tourist entering the VO requests the RouteDesign service through his agent. Then the RouteDesign service collects the properties of the tourist through Metadata service in step 2, 3. Next, in step 4 RouteDesign service generates an Authorization request which contains the RDF based property regarding consuming points. In step 5 and 6, Authorization service gets VO ontology containing the role definitions from the ontology service. And in step 7 and 8, Authorization service invokes the Reasoning service to infer the role of the tourist by passing the VO ontology and the tourist property as parameters. Then Authorization service compares the inferred tourist's roles and the role based access control list to evaluate the eligibility of the tourist in step 9. Finally, if the access is allowed, the RouteDesign service is invoked in step 10 and returns the travel route to tourist agent in step 11. If denied, no route computing is executed and the deny information is returned to the tourist agent in step 11.

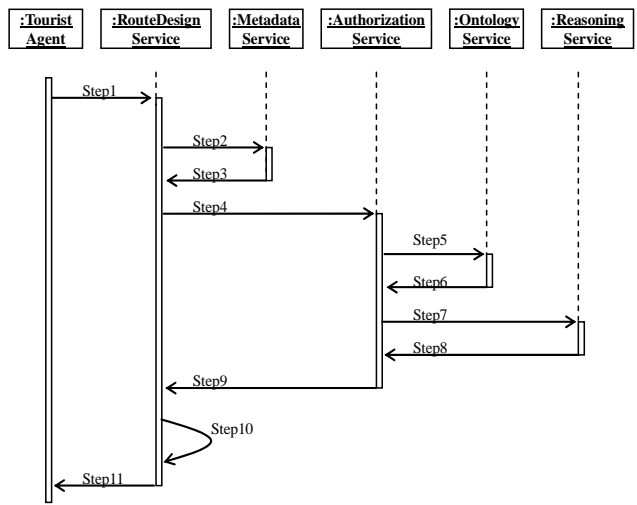

**Fig. 7.** Authorization Process Based on Roles

### 3.6 Intelligent Agent

From the scenario in fig. 1, there is a clear need of agents to perform some flexible, autonomous actions to accomplish e-Tourism tasks [25]. Some agents act as the representative of the tourists or travel agencies to forward the requests, monitor the status and receive the responses. While some other agents are capable of mining preference, which aids in precise and timely targeting, and personalization of tourism products. To achieve this, the agent need to trace the user's requests, analyze his preferences, and continually refresh the dynamic part of user profile like interests. In the following example, an Interests-Value array $IV_{ij}$ ($i=1...n$ items of interests, $j=0, 1$) is used by the agent to represent relevant tourist's interests.

when $j=0$, $IV_{ij}=\{items\ of\ interests\}$;
when $j=1$, $IV_{ij}=\{1,2,3,4,5\}$.   /*value of interests:
        1, invokes of service in a month $<5$;
        2, invokes of service in a month $<10$ and $>=5$;
        3, invokes of service in a month $<15$ and $>=10$;

*4, invokes of service in a month <20 and >=15;*
*5, invokes of service in a month >=20 */*

From the above definition, *items of interests* is a set of requests invoked by the tourist in the past month. *Value of interests* reflects the frequency of requests invokes by the tourist in the past month. More frequently a request is invoked by a user, the higher the *value of interests* is. If the *value of interests* of a request, for example "Request (op: airFare, source: Beijing, destination: Newyork )", is greater than a threshold, then this request will be added to a preference list and invoked by the agent periodically, thus the updated air fare from Beijing to Newyork will be automatically returned to the tourist.

For complex e-Tourism business, the agent need to be enhanced by workflow and rule based reasoning engine. Workflow engine enables multiple agents to collaborate in a customizable way, while reasoning engine can assign agents the capacity to infer knowledge through declarative rules. There have been lots of discussions on reasoning and workflow of agents[26] [27] [28] [29].

## 4    Implementation

The framework is implemented on two layers: application and infrastructure. Application layer contains semantic aware e-Tourism grid services (e.g. RouteDesign service), which solve the business problems in e-Tourism domain and are deployed in the containers supported by the lower semantic grid infrastructure. The infrastructure layer conforms to the S-OGSA and is implemented by OntoGrid framework [30]. It deploys Globus Toolkit 4 as basic grid platform, integrates Jena [31] to support semantic Web functionalities (for example, RDF based metadata storage and reasoning), and uses Apache Axis [32] and WSRF [33] to declare Web services and execute SOAP requests. Moreover, several customizations and enhancements like ontologies, Mediator and intelligent agent have been applied on OntoGrid framework to better serve the e-Tourism requirements. Specially, the agent facilities are set up through JADE (Java Agent Development Evnironment) [34] with extensions of reasoning and workflow engines: JESS (Java Expert System Shell) [29] and Wade (Workflow and Agent Development Environment) [26]. Additionally, integration between agent facilities and S-OGSA framework is completed through a service oriented manner in which the communication is realized via Web services.

## 5    Related Work

The semantic Web and semantic Web service has substantial effects on next generation e-Tourism infrastructure, which enhances existing tourism Web resources and services through semantic way, makes them "smarter" and capable of carrying out intelligent reasoning behind the scenes. Some typical projects include:
- Harmonise [35] is an EU Tourism Harmonisation Network established by eCTRL, IFITT and others. It creates an electronic space for tourism stakeholders to exchange information in a seamless, semiautomatic manner, independent from geographical, linguistic and technological boundaries. It defines an Interoperability Minimum Harmonisation Ontology for modeling and saving concepts of transaction data.
- Hi-Touch project [36] is to develop semantic Web methodologies and tools for intra-European sustainable tourism. It makes use of the WTO Thesaurus on Tourism and Leisure Activities as an authoritative source for its ontology.
- OnTour project [37] developed by e-Tourism Working Group at Digital Enterprise Research Institute, designs an advanced e-Tourism semantic Web portal connecting the customers and virtual travel agents, and an e-Tourism ontology using OWL and WTO Thesaurus.
- SATINE project [13] realizes a semantic based infrastructure which allows the Web services on well-established service registries like UDDI or ebXML to seamlessly interoperate with Web services on P2P Networks. The travel ontologies are developed based on standard specifications of Open Travel Alliance (OTA).

Despite contributing a lot for the e-Tourism infrastructure, semantic Web and semantic Web services have the weakness in flexible computation, storage, VO and collaboration which are critical to next generation e-Tourism. Semantic grid, however, is able to compensate these lacks by seamlessly integrating grid facilities with semantic Web and Web services. Currently several pioneering applications based on semantic grid have been developed in the following:
- GRISINO project aims to develop an experimental test-bed combining advanced prototypes of each of the three technologies: Knowledge Content Objects as a model for the unit of value, WSMO/L/X as a framework for the description and execution of semantic Web services and Globus as the grid infrastructure for managing resources and hosting services.
- UK e-Science program has reinforced the practical need for the semantic grid, and funded many e-Science projects based on various semantic grid solutions, like CombeChem/eBank, CoAKTinG, MIAKT and Medical Devices.
- There are still more significant efforts to provide an architecture like S-OGSA for the development of semantic grid applications or simply semantic aware grid Services, such as projects InteliGrid [38] and myGrid, showing how explicit metadata can be used in the context of existing grid applications.

## 6   Conclusion

This paper has discussed the semantic grid as the next generation e-Tourism infrastructure, which supports high degree automation, seamless integration, knowledge services, intelligent agent, flexible collaboration, and sharing of computation and storage resources on a global scale. It's not difficult to imagine a lot of innovative e-Tourism applications on this promising infrastructure. However, the semantic grid is still in its early experimentation phase of pioneering applications and far away from the mature [23]. To make it a reality, there are still many research challenges, for example, performance, scalability, reliability and security problems.

# References

1. Cardoso, J.: E-Tourism: Creating Dynamic Packages using Semantic Web Processes. In: W3C Workshop on Frameworks for Semantics in Web Services. Innsbruck(2005)
2. Keun, H.K., Jeong, S.H., Pilsoo, S.K.: Modeling for Intelligent Tourism E-Marketplace Based on Ontology. In: Proc. of the 2007 International Conference on Recreation, Tourism, and Hospitality Industry Trends, pp. 56-65. Taiwan (2007)
3. Berners-Lee, T., Hendler, J., Lassila, O.: Semantic Web. J. Scientific American. 284, 5, pp. 34–43 (2001)
4. Murphy, M.J., Dick, M., Fischer, T.: Towards the Semantic Grid. J. Communications of the IIMA. Vol. 8, Issue 3, pp. 11-24 (2008)
5. Roure, D.D.: Future for European Grids: GRIDs and Service Oriented Knowledge Utilities. In: Vision and Research Directions 2010 and Beyond, http://www.semanticgrid.org/documents/ngg3/ngg3.html
6. Zhuge, H.: Semantic Grid: Scientific Issues, Infrastructure, and Methodology. J. COMMUNICATIONS OF THE ACM. vol. 48, No. 4, pp. 117-119 (2005)
7. Corcho, O., Alper, P., Kotsiopoulos, I., Missier, P., Bechhofer, S., Goble, C.: An overview of S-OGSA: A Reference Semantic Grid Architecture. J. Web Semantics. vol. 4, pp. 102-115 (2006)
8. Roure, D.D., Jennings, N.R., Shadbolt, N.R.: The Semantic Grid: A Future e-Science Infrastructure. In: Grid Computing - Making the Global Infrastructure a Reality, pp. 437-470. John Wiley and Sons Ltd (2003)
9. Foster, I., Kishimoto, H., Savva, A., Berry, D., Djaoui, A., Grimshaw, A., Horn, B., Maciel, F., Siebenlist, F., Subramaniam, R., Treadwell, J., Reich, J. V.: The Open Grid Services Architecture, Version 1.0. Technical report, Global Grid Forum (2005)
10. Foster, I.: A Globus Toolkit Primer, www.globus.org/primer (2005)
11. Prantner, K., Ding, Y., Luger, M., Yan, Z., Herzog, C.: Tourism Ontology and Semantic Management System: State-of-the-arts Analysis. In: IADIS International Conference WWW/Internet 2007. Vila Real, Portugal (2007)
12. Maedche, A., Motik, D., Silva, N., Volz, R.: MAFRA-A MApping FRAmework for Distributed Ontologies. In: Proc. of the 13th European Conf. on Knowledge Engineering and Knowledge Management EKAW-2002. Madrid, Spain (2002)
13. Dogac, A., Kabak, Y., Laleci, G., Sinir, S., Yildiz, A., Tumer, A.: SATINE Project: Exploiting Web Services in the Travel Industry. In: eChallenges 2004. Vienna (2004)
14. Dogac, A., Kabak, Y., Laleci, G., Sinir, S., Yildiz, A., Kirbas, S., Gurcan, Y.: Semantically Enriched Web Services for the Travel Industry. J. ACM Sigmod Record, Vol. 33, No. 3 (2004)
15. Zhang, X.M: High performance virtual distributed object. J. Journal of Computer Research and Development. Supplement, pp. 102-107 (2000)
16. Zhang, X.M: A Dynamic Scalable Asynchronous Message Model Based on Distributed Objects. J. Computer Engineering and Science, vol. 3, pp. 48-50 (2002)
17. GT 4.0 WS MDS Index Service: System Administrator's Guide, http://www-unix.globus.org/toolkit/docs/development/4.0-drafts/info/index/admin/
18. S-MDS: semantic monitoring and discovery system for the Grid. J. Grid Computing, vol. 7, pp. 205–224 (2009)
19. Sharman, N., Alpdemir, N., Ferris, J., Greenwood, M., Li, P., Wroe, C.: The myGrid Information Model. In: UK e-Science programme All Hands Conference (2004)
20. myGrid: The myGrid project, http://www.mygrid.org.uk/ (2008)
21. Miles, S., Papay, J., Payne, T.R., Decker, K., Moreau, L.: Towards a protocol for the attachment of semantic descriptions to Grid services. In: European Across Grids Conference, pp. 230–239 (2004)

22. Lee, F., Garg, S., Garg, S.: OWL-SR: Unified Semantic Service and Resource Discovery for Grids. In: The 4th European Semantic Web Conference (2007)
23. Alper, P., Corcho, O., Parkin, M., Kotsiopoulos, I., Missier, P., Bechhofer, S., Goble, C.: An authorisation scenario for S-OGSA. In: Proceedings of Posters and Demos, 3rd European Semantic Web Conference, ESWC06, pp. 7–8 (2006)
24. Brooke, J.M., Parkin, M.S.: Enabling scientific collaboration on the Grid. J. Future Generation Computer Systems (2008)
25. Wooldridge, M.: Agent-based software engineering. In: IEE Proc on Software Engineering, vol. 144 (1), pp. 26-37 (1997)
26. Caire, G., Gotta, D., Banzi, M.: WADE: a software platform to develop mission critical applications exploiting agents and workflows. In: Proc. of the 7th international joint conference on Autonomous agents and multiagent systems. pp. 29-36. Estoril (2008)
27. Buhler, P.A., Vidal, J.M.: Towards Adaptive Workflow Enactment Using Multiagent Systems. J. Information Technology and Management. Vol. 6(1), pp. 61-87 (2005)
28. Negri, A., Poggi, A., Tomaiuolo, M.: Dynamic Grid Tasks Composition and Distribution through Agents. J. Concurrency and Computation, vol. 18(8), pp. 875-885 (2006)
29. JESS: the Rule Engine for the JavaTM Platform, http://herzberg.ca.sandia.gov/jess/ (2009)
30. Goble, C., Kotsiopoulos, I., Corcho, O., Missier, P., Alper, P., Bechhofer, S.: S-OGSA as a Reference Architecture for OntoGrid and for the Semantic Grid. In: GGF16 Semantic Grid Workshop (2006)
31. Carroll, J.J., Dickinson, I., Dollin, C., Reynolds, D., Seaborne, A., Wilkinson, K.: Jena: implementing the semantic web recommendations. In: Proc. of the 13th international World Wide Web conference. New York (2004)
32. Axis Architecture Guide, http://ws.apache.org/axis/java/architecture-guide.html
33. Czajkowski, K., Ferguson, D., Foster, I., Frey, J., Graham, S., Sedukhin, I., Snelling, D., Tuecke, S., Vambenepe, W.: Web Services Resource Framework (WSRF). Technical report, Globus Alliance and IBM (2005)
34. Bellifemine, F., Poggi, A., Rimassa, G.: Jade: a fipa2000 compliant agent development environment. In: Proceedings of the fifth international conference on Autonomous agents, pp. 216-217. ACM Press (2001)
35. Missikoff M., Werthner H., Hopken W., et al.: Harmonise-Towards Interoperability in the Tourism Domain. In: Proc. of the 10th International Conference on the Information and Communication Technologies in Travel & Tourism. Helsinki, Finland, (2003)
36. Hi-Touch project, http://icadc.cordis.lu/fepcgi/srchidadb ? CALLER=PROJ_IST & ACTION=D & RCN=63604 & DOC=20 & QUERY=3
37. Bachlechner, D.: OnTour - The Semantic Web and its Benefits for the Tourism Industry, http://e-tourism.deri.at/ont (2005)
38. Dolenc, M., Turk, Ž., Katranuschkov, P., Krzysztof, K.: D93.2 Final report of the InteliGrid. Technical report, The InteliGrid Consortium and University of Ljubljana (2007)